\newcommand{\version}{June 18, 2008}

\documentclass[12pt]{article}
\usepackage{bbm}
\usepackage{a4,amsthm,amsfonts,latexsym,amssymb}
\usepackage{graphicx}

\swapnumbers 

\theoremstyle{plain}
\newtheorem{thm}{THEOREM}

\newtheorem{lem}[thm]{LEMMA}
\newtheorem{define}[thm]{DEFINITION}
\newtheorem{proposition}[thm]{PROPOSITION}

\newcommand{\beq}{\begin{equation}}
\newcommand{\eeq}{\end{equation}}
\def\beqa{\begin{eqnarray}}
\def\eeqa{\end{eqnarray}}

\newcommand{\tab}{\hspace*{7mm}}
\newcommand{\ifa}{$\Leftrightarrow$}

\newcommand{\C}{{\mathbb C}}

\newcommand{\R}{{\mathbb R}}

\newcommand{\s}{{\mathbf{S}}}
\newcommand{\one}{{\mathbbm 1}}
\newcommand{\Tr}{{\rm Tr}}
\newcommand{\rank}{{\rm rank}}
\newcommand{\range}{{\rm range}}

\newcommand{\B}{{\mathcal B}}
\newcommand{\G}{{\mathcal G}}
\newcommand{\Hh}{{\mathcal H}}
\newcommand{\Nn}{{\mathcal N}}
\newcommand{\K}{{\mathcal K}}

\newcommand{\F}{{\mathcal F}}

\newcommand{\D}{{\mathcal D}}
\newcommand{\TT}{{\mathcal T}}

\newcommand{\eps}{\varepsilon}
\newcommand{\lda}{\lambda}
\newcommand{\E}{{\mathcal E}}

\newcommand{\half}{\mbox{$\frac{1}{2}$}}
\newcommand{\der}{\mbox{$\frac{d}{dt}$}}

\def\bce{\begin{center}}
\def\ece{\end{center}}
\def\bit{\begin{itemize}}
\def\eit{\end{itemize}}

\date{\small\version}

\begin{document}
\markboth{\scriptsize{SGL2 BN  \version}}{\scriptsize{SGL2 BN \version}}

\title{
\vspace{-80pt}
\begin{flushright}
{\small Vienna, ESI Report 2034 (2008)} \vspace{30pt}
\end{flushright}
\bf{Analysis of quantum semigroups with GKS--Lindblad generators II.
\newline
General}}
\author{\vspace{8pt} Bernhard Baumgartner$^1$, Heide Narnhofer$^2$
\\
\vspace{-4pt}\small{Fakult\"at f\"ur Physik, Universit\"at Wien}\\
\small{Boltzmanngasse 5, A-1090 Vienna, Austria}}

\maketitle


\begin{abstract}
Semigroups describing the time evolution of open quantum systems in finite-dimensional spaces have
generators of a special form, known as Lindblad generators. These generators and the corresponding
processes of time evolution are analyzed,
characterized as Decay, Dissipation and Dephasing.
In relation to these processes the Hilbert space of the system is equipped with a special structure,
a decomposition into a sum of mutually orthogonal subspaces.
The complete set of all the stationary states and the asymptotic behavior
of the evolutions are presented in detail.
Some unusual special facts about invariant operators and symmetries
are studied, examples are demonstrated.
Perturbation theory for the structure and for the stationary states is discussed
and performed in case studies.
\\[20ex]
PACS numbers: \qquad  03.65.Yz , \quad 05.40.-a , \quad 42.50.Dv ,
\quad 03.65.Fd
\\[3ex]
Keywords: open system, time evolution, Lindblad generator, semigroup

\end{abstract}

\footnotetext[1]{\texttt{Bernhard.Baumgartner@univie.ac.at}}
\footnotetext[2]{\texttt{Heide.Narnhofer@univie.ac.at}}


\newpage

\section{Introduction}\label{intro}

Hundreds of papers dealing with ``Lindblad equations''
have been written since their fundamental importance in the theory of open quantum
systems was shown (see \cite{L76, GKS76}). They give the
proper mathematical form to Markovian semigroups of completely positive norm continuous maps
which are needed in this context \cite{D76}.
The early investigations (see, for example \cite{S76,F78,S80}) mostly
aimed at establishing theories for the approach to thermal equilibrium,
and all the subsequent investigations pursued special physical questions;
none of them, however, dealt with the general mathematical structure.
And this is now the theme of this paper.
We give an analysis of the mathematical properties. It is analogous to the functional
analysis of unitary groups of evolution, which was presented in the early time of quantum mechanics.
We continue the study on semigroups of completely positive maps, acting on states
for systems with finite dimensional Hilbert spaces, which we began in \cite{BNT08}.

While Schr\"{o}dinger equations are studied in Hilbert space spanned by the pure-state vectors,
the Lindblad equations will be studied on two levels: One is the linear space of density matrices,
the other is the underlying Hilbert space.
A relation between these two levels is established.
In this relation we find a clear distinction between different processes:
Decay, Dissipation and Dephasing.
Characterizing these processes involves a structuring of the underlying Hilbert space $\Hh$.
Each Lindblad generator is related to a special decomposition of $\Hh$ into mutually orthogonal subspaces.
This decomposition can be seen as a generalization of the spectral decomposition of Schr\"{o}dinger operators.
Orthogonality of the subspaces holds in spite of non-hermiticity of the defining operators.

We do not discuss ``decoherence''. The meaning of this concept in physics is connected with
the transition from quantumness to classicality, \cite{J03, Z03a, Z03b}, the mathematics of it requires the
definition of a preferred basis. ``Dephasing'' is a similar concept,
and we define what it means precisely in a mathematical fashion.

Our starting point, the connection with the earlier studies,
is the result of \cite {GKS76, L76}:
\begin{proposition} {\bf Generators of semigroups:}\label{lindbladprop}
Every generator of a semigroup of completely positive trace preserving maps
$\TT^t:\, \rho(s)\mapsto\rho(s+t)$ for $t\geq 0$
on the set of finite dimensional density matrices $\rho$, can be written in the form
\beq\label{lindblad}\nonumber
\dot{\rho}=\D(\rho)= -i[H,\rho]  +\sum_\alpha \D_{h_\alpha}(\rho)
\eeq

where  $H=H^\dag$ is a hamiltonian. Transition operators $h_\alpha$ define the irreversible parts, the
{\bf simple generators:}
\beq\label{simplegenerator}
\D_h (\rho)=h\rho h^\dag -\half (h^\dag h\rho+\rho h^\dag h)
\eeq
\end{proposition}
This representation of $\D$ is known as ``diagonal''.
For our purposes it is optimal. All what is needed to characterize the processes
can be extracted from the set of the operators $\{H, h_\alpha, h_\alpha^\dag\}$,
and from the algebra of operators which commute with each element of this set.
(A remark on the notation: We use here $h_\alpha$ with lower index,
instead of the use of an upper index, which has been employed in our first paper \cite{BNT08}.)
It is well known, \cite{AF01, BP02}, that the division of $\D$ into a sum of several simple generators,
attribution of $\{H, h_\alpha\}$ to $\D$, is not unique.
Different sets of operators $h_\alpha$ can be attributed to a given generator $\D$,
some actions can be shifted from the Hamiltonian $H$ to the transfer operators $h_\alpha$
or in the other way.
There exist attempts to favor special ways, like demanding $\Tr[h_\alpha ]=0$,
but we make here, in this paper, no restriction.

The mathematical analysis is of course related to special questions concerning physics:
\begin{itemize}
\item  Existence and characterization of stationary states.
\item Geometry of the paths and characterizing Hilbert space subspaces.
\item Aspects of symmetries.
\item Perturbation.
\end{itemize}

Identifying stationary states is closely related to identifying
special subspaces of the Hilbert space.
The summary of our investigations on the first two of these themes can be stated as follows:

\begin{thm} {\bf Structuring of the Hilbert space:}\label{structurthm}
\begin{enumerate}
\item {\bf Decay: }
The Hilbert space can be represented in a unique way as a direct sum of two orthogonal subspaces,
$\Hh=P_0\Hh\oplus P_0^\perp\Hh$,
where $P_0^\perp\Hh$ is the maximal decaying subspace, i.e.
\beq
\forall \rho\in\s :\quad \lim_{t\to\infty}P_0^\perp\TT^t(\rho)P_0^\perp=0,
\eeq
and $P_0\Hh$ contains no decaying sub-subspace.
\beq
\forall Q\leq P_0\,\ \exists \rho\in\s :\quad \limsup_{t\to\infty}\Tr[Q\TT^t(\rho)Q]\neq 0.
\eeq

\item {\bf Dephasing: }
The collecting subspace  $P_0\Hh$ can further be divided in a unique way by splitting
$P_0=\sum_k Q_{0,k}$,\,\, $Q_{0,k}\cdot Q_{0,\ell}=\delta_{k\ell}Q_{0,k}$,
into minimal subspaces $Q_{0,k}\Hh$ with relative dephasing
\beq
\forall\rho(0),\quad\forall k, \ell, \quad k\neq\ell:\quad
\lim_{t\to\infty}Q_{0,k}\rho(t)Q_{0,\ell}=0,
\eeq
and the property, that the time evolution of each block $Q_{0,k}\rho(t)Q_{0,\ell}$
is independent of the other blocks.

\item {\bf Asymptotics: }
Each subspace $Q_{0,k}\Hh$ can be represented as $Q_{0,k}\Hh = \C^{n(k)}\otimes \Hh_{00,k}$,
so that the time evolution at large times is described in block form, with Hamiltonians
$H_{0,k}\otimes\one$ acting on $Q_{0,k}\Hh$ and with unique density matrices acting on $\rho_k$ on $\Hh_{00,k}$
inside each block:
\beq
\forall\rho(0)\quad\exists\,\,\{\lambda_k,\, R_k\}:\quad
\lim_{t\to\infty}|\rho(t)-\bigoplus_k \lambda_k e^{-iH_{0,k} t} R_k e^{iH_{0,k} t}\otimes \rho_k|=0,
\eeq
where each $R_k$ is a positive matrix with trace one acting on $\C^{n(k)}$,
and $0\leq \lambda_k\leq 1$, $\sum_k\lambda_k=1$.
\\The set of stationary states is given by the set of density matrices
\beq
\bigoplus_k \lambda_k R_k\otimes \rho_k ,
\eeq
where $[H_{0,k},R_k]=0$, and $\rho_k$ again unique for each $k$.

\item {\bf Dissipation: }
The limiting density matrices inside the minimal blocks are of
maximal rank: $\rank(\rho_k)=\dim (\Hh_{00,k})$.

\item {\bf Cascades with basins: }
The decaying subspace $P_0^\perp\Hh$ can further be divided by splitting it
as a cascade with ``basins'' $P_{k,\ell}\Hh$, $k\geq 1$,
all mutually orthogonal, $P_{k,\ell}\cdot P_{j,m}=\delta_{kj}\delta_{m\ell}P_{k,\ell}$,
and arranged in levels $P_k\Hh=\bigoplus_\ell P_{k,\ell}\Hh$.
So $P_0^\perp=\sum_{k\geq 1,\ell} P_{k,\ell}$, and
the time evolution is like a ``flow'', out of each basin into the ``lower'' levels
including collecting basins $P_{0,k}\Hh\in P_0\Hh$, where each $Q_{0,\ell}\Hh$
mentioned in the item above contains $n(\ell)$ basins:
\beqa
\textrm{for}\quad t>0:\quad \quad\quad
P_j\cdot\TT^t (P_{k,\ell}\rho P_{k,\ell})\cdot P_j&\neq& 0\quad\quad \textrm{if}\quad j<k,\\
P_j\cdot\TT^t (P_{k,\ell}\rho P_{k,\ell})\cdot P_j&=& 0\quad\quad \textrm{if}\quad j>k,\\
P_{k,m}\cdot\TT^t (P_{k,\ell}\rho P_{k,\ell})\cdot P_{k,m}&=& 0\quad\quad \textrm{if}\quad m\neq \ell.
\eeqa
In special cases there is a possibility of unitarily reshuffling some basins, defining other basins
$\tilde{P}_{k,\ell}=U\cdot P_{k,\ell}\cdot U^\dag$.
The number of minimal basins is unique, as are their dimensions.
\end{enumerate}
All projectors $P_k, P_{k,\ell}$ and $Q_{0,k}$ are orthogonal projectors.
\end{thm}

Perpendicular to the structuring into levels there are dissections into ``enclosures'' with
a division of density matrices into blocks with mutually independent evolutions.
\begin{thm} {\bf Enclosures and blocks:}\label{conventional}
If there exists a set of mutually orthogonal projectors $Q_m$, each commuting with $H$ and every $h_\alpha$,
then the basins $P_{k,\ell}\Hh$ can be chosen in such a way that each subspace $Q_m\Hh$
is a direct sum of basins. We call such a $Q_m\Hh$ an enclosure.
The time evolution of any density matrix $\rho$
splits into mutually independent evolutions of blocks $Q_m\rho Q_\ell$:
\beq
\TT^t(Q_m\rho Q_\ell)=Q_m \TT^t(\rho) Q_\ell
\eeq
\end{thm}

As an example for the structuring, already well known, may serve the Grotrian diagram for Helium.
The dissection into enclosures is there the distinction between ortho- and para-helium.
The lowest energy levels represent the collecting subspace.
Energy levels with angular momentum zero are basins,
those with higher angular momenta $\vec{J}$ can further be split into basins,
by diagonalizing a component of $\vec{J}$.
The choice of the axis of the component is not unique,
but different kinds of splitting, with different basins, are unitarily equivalent.

The $Q_m$ are projectors onto the ``enclosures'' -- for details see Section \ref{definitions}.
In the Heisenberg picture, with time evolution of the observables,
they are invariant.
This is a conventional symmetry.
But there are cases when other symmetries turn up in non conventional way, without conserved observables;
cases when conservation of observables appears, without an overall symmetry;
and the conserved observables need not commute with all the $h_\alpha$, and they need not form an algebra.
``Degeneracy'', the existence of several stationary states, may be connected with
occurrence of symmetries or not.
The well known connection of symmetries
with conservation laws, fundamental for Langrangian mechanics and quantum mechanics,
is here no longer valid. All this is discussed in Section \ref{symmetry}.
It is also an important aspect when perturbations
are studied, as is done in Section \ref{perturbation}.

\section{Basic properties of the superoperators}\label{super}

\subsection{The Hilbert Schmidt space of matrices}\label{hsmatrices}

Consider the set $\s$ of states as the set of $n\times n$ density matrices
$\rho_{i,j}=\langle i|\rho|j\rangle$, with $n=\dim(\Hh)$, $\{|i\rangle\}$ some basis of $\Hh$,
and consider it as embedded into the linear space of complex $n\times n$ matrices.
Now consider the general mathematical wisdom on
linear differential equations with constant coefficients, acting in a finite dimensional space,
which is here the space spanned by $n\times n$ dimensional matrices:
To each superoperator $\D$
there exist eigenmatrices $\sigma$, proper or generalized:
$$\D(\sigma)=\lambda \sigma\quad \textrm{or} \quad(\D-\lambda)^n(\sigma)=0,$$
So one has the special time evolutions:
$\sigma(t)=e^{\lambda t}\cdot Polynomial(t, \D^n(\sigma))$.
The general solution to $\dot{\sigma}=\D(\sigma)$ is a linear combination of these special solutions.
But the eigenmatrices to our $\D$ may be not self-adjoint,
although $\D$ preserves self-adjointness. Instead, there may be pairs of  eigenvalues
$\{ \lambda,\lambda^\ast\}$
 with pairs of eigenmatrices $\{ \sigma,\sigma^\dag\}$.
That's the reason why we do not stay in the space of self-adjoint matrices.

We endow this linear space with the Hilbert Schmidt norm
$\|\sigma\|=(\Tr [\sigma^\dag\sigma])^{\half}$ and the inner product
$\langle\langle\sigma|\rho\rangle\rangle =\Tr [\sigma^\dag\rho]$.
The HS-superspace of operators enables the definition of the adjoint of a super-operator.
The adjoint super-operator $\D^\dag$
generates the evolution of observables in the Heisenberg-picture, $\dot{F}=\D^\dag(F)$.
It acts as

\beq\label{heisenberg}
\D^\dag(F)= i[H,F]  +\sum_\alpha \D_{h_\alpha}^\dag(F),
\eeq
where
\beq\label{heisenbergsimple}
\D^\dag_h (F)=h^\dag F h -\half (h^\dag h F+ F h^\dag h).
\eeq

We remark that self-adjointness of $\D$ is rather exceptional.
See Section \ref{examples} for examples.

\subsection{Evolutions inside the set of states}\label{evolutioninside}

$\TT^t$ preserves self-adjointness, trace-norm and, for $t \geq 0$, positivity;
it maps $\s\mapsto \s$, \,\, the set of normed positive matrices into itself.

The dissipative character of the evolution is reflected in the general
changes of eigenvalues $r(t)$ of $\rho(t)$.
Zero eigenvalues may become positive, positive eigenvalues may increase or decrease. But
the decrease is bounded from below:
\begin{proposition}{\bf A differential inequality:}\label{inequality}
If $r(t)$ is a non-negative eigenvalue of $\rho(t)$, its change in time is bounded from below
as
\beq
\dot{r}(t)\geq -\left(\sum_\alpha \| h_\alpha\|^2\right)r(t).
\eeq
\end{proposition}
\begin{proof}
Differentiating the eigenvalue equation
$\rho(t)\psi(t)=r(t)\psi(t)$,
and using $\langle\psi|\dot{\psi}\rangle=0$
gives \beqa
\dot{r}(t)=\langle\psi|\dot{\rho}|\psi\rangle &=&
\sum_\alpha\langle\psi|(h_\alpha \rho h_\alpha^\dag -
\half(h_\alpha^\dag h_\alpha\rho+\rho h_\alpha^\dag h_\alpha))|\psi\rangle \nonumber\\
&\geq& -\sum_\alpha\half\langle\psi|(h_\alpha^\dag h_\alpha\rho+\rho h_\alpha^\dag h_\alpha)|\psi\rangle \nonumber\\
&=&-\sum_\alpha\langle\psi|h_\alpha^\dag h_\alpha|\psi\rangle r(t).
\eeqa
Here, the finite dimension of the Hilbert space is essential: it implies that
the number of simple generators which are necessary to form some given $\D$ is bounded by
$\dim^2(\Hh)-1$. All the sums over $\alpha$ are finite.
\end{proof}
An immediate consequence is that
the positive eigenvalues $r_j$ of the density operator obey the inequality
$$r_j(t)\geq\exp\left(-\sum_\alpha \| h_\alpha\|^2 t\right)\cdot r_j(0)>0.$$
The rank of $\rho$ cannot decrease in finite time,
purification can only occur in the limit $t\to\infty$.

Moreover, we can affirm quite generally
\begin{proposition}{\bf Existence of stationary states.}\label{existence}
For each $\D$ there exists at least one stationary state $\rho\in\s$.
\end{proposition}
\begin{proof}
The density matrices in $\B(\Hh)$ form a compact convex set $\s$.
We know that the semigroup maps $\s$ into itself,
and we may consider the map
$\rho(0) \mapsto \rho(\Delta t)$
for some fixed time interval $\Delta t$.
By the Brouwer fixed point theorem, there exists at least one fixed point,
$\rho(\Delta t) = \rho(0)$.
So there exists the stationary state
\beq
\bar{\rho}= \frac1{\Delta t} \int_0^{\Delta t}\rho(t)dt,
\eeq
satisfying $\D(\bar{\rho})=0$.
\end{proof}
This abstract statement is needed in \ref{dissipation}. In the following
it is then elucidated by more concrete formulas.

\section{Processes and Structuring}\label{structuring}

\subsection{Definitions}\label{definitions}

\textbf{Enclosure} is not a standard concept, although it appears in Hamiltonian dynamics
as a consequence of the conservation of energy.
We define ``enclosure'' as a subspace  $\K$ with the property
that for all $\rho(0)$ the expectation of the orthogonal projector onto
this subspace, $Tr[P_\K\rho(t)]$, is constant in time.
In the Schr\"{o}dinger picture this means that the system can neither leave
nor enter this subspace. In the Heisenberg picture it means invariance of the observable
$P_\K$ in the course of time. Each $P_\K$ is one of the projectors $Q_m$ appearing in Theorem \ref{conventional}.
So, enclosure is a case of ``conservation'', but in the special way that it is an orthogonal projector
which is the conserved observable. In Section \ref{examples} it is shown that there
are cases where invariant observables exist, but no invariant projector,
different to the evolutions under Hamiltonian dynamics.

Our analysis of general Lindblad equations $\TT^t$ starts with investigating
an eventual enclosing of the system in the subspaces $\K$ with conserved projectors $P_\K$.
Then, \emph{inside} the enclosures, decay and/or dissipation occurs.
The phase-relations between the enclosures may eventually show a dephasing.
Dephasing is a typical quantum-effect, but it is also related to \emph{decoherence},
the transition from quantum to classical appearance, \cite{Z03a,Z03b}.
We introduce precise mathematical definitions of these physical events
of irreversible processes:
\begin{define}{\bf Irreversible events:}
\\\textbf{Decay} occurs, if there exists a decaying subspace $\K$; i.e.
\beq
\forall\rho(0):\,\,\Tr[P_\K\rho(t)]\rightarrow 0.
\eeq
\textbf{Dissipation} occurs, if there exists a subspace $\K$ such that
\beq
\forall\rho(0)=P_\K\rho(0): \, \rank(\rho(t))\rightarrow dim(\K).
\eeq
\textbf{Dephasing} occurs for phase relations, in the case of conservations of
$P_\K$ and $P_\mathcal{L}$, with $\K\perp\mathcal{L}$,
if
\beq\forall\rho(0):\,\,P_\K\,\rho(t)\,P_\mathcal{L}\rightarrow 0.
\eeq
All arrows are meant as limit of $t$ to infinity.
\end{define}

Enclosures, decaying and - their complement - collecting subspaces are the analogue to the energy eigenspaces in
Hamiltonian dynamics.

\subsection{Enclosures and conservation of projectors}\label{consevation}

In the Heisenberg picture ``conservation'' of $P$ means $\D^\dag(P)=0$.
In the Schr\"{o}dinger picture this means that the state of the system can neither leave
nor enter the subspace $P\Hh$.
\begin{lem}\label{vneumann}
Conservation of $P=P^\perp=P^2$ is possible if and only if this observable commutes both with
the Hamiltonian $H$ and with all the transfer operators $h_\alpha$; in other words, iff
it is an element of the von Neumann algebra $\{H,h_\alpha\}'$.
\end{lem}
\begin{proof}
We consider $\rho=P\rho P$, systems in the subspace $P\Hh$.
Leaving this subspace is possible for the system, if
$$P^\perp\D(\rho)P^\perp=\sum_\alpha P^\perp h_\alpha\,\rho\, h_\alpha^\dag P^\perp\neq 0\quad\Leftrightarrow\quad
\exists\alpha: \quad P^\perp h_\alpha P \neq 0.$$
Entering the same subspace means leaving the orthogonal complement and is thus possible for the system if
$\exists\alpha:\,\, P h_\alpha P^\perp \neq 0$.
That neither leaving nor entering is possible implies therefore $\forall\alpha:\,\, Ph_\alpha=Ph_\alpha P=h_\alpha P$.
Under this restriction one gets for the evolution of the phase relations,
the off diagonal part, $P\,\rho(t)\,P^\perp$:
$$P\D(\rho)P^\perp=-iP[H,\rho]P^\perp=iP\,\rho\, P\,H\,P^\perp,$$
which vanishes for every $\rho=P\rho P$ iff $H$ commutes with $P$.
And its vanishing is necessary, otherwise
$P^\perp \D^2(\rho)P^\perp = P^\perp H P\rho P H P^\perp>0$.
\end{proof}

Assume that both $P_i$ and $P_j$ are conserved projectors.
The commutations imply $\D(P_i\rho)=P_i\D(\rho)$ and $\D(\rho P_j)=\D(\rho)P_j$.
So the evolution of each block $P_i\rho P_j$ is independent from the other parts of $\rho$.
The set of conserved projectors generates the algebra $\Nn=\{H,h_\alpha\}^\prime$.
A maximal abelian subalgebra of the von Neumann algebra $\Nn$ gives
a set of mutually orthogonal minimal conserved projectors $P_i$,
a decomposition of $\Hh$, and a decomposition of $\rho$ as a block-matrix,
with split evolutions of the blocks.
Such a decomposition is unique if $\Nn$ is an abelian algebra.
If this algebra is not abelian, then different decompositions are possible.
They are related by unitary transformations with $U\in\Nn$ which leave $\D$ invariant.
This follows from the classification of finite dimensional von Neumann algebras, stated
for example in \cite{J03b}.

Conserved projectors come along with an algebra of conserved observables, $\{H,h_\alpha\}^\prime$,
and, moreover, with a dynamical symmetry:
\beq
\TT^t(U^\dag\rho U)=U^\dag\rho(t)U\quad \textrm{if} \quad U\in\{H,h_\alpha\}^\prime.
\eeq
But there are cases of conserved observables which do not form an algebra;
and, on the other hand, conservation of projectors is not necessary for occurrence of a dynamical symmetry.
This is discussed in Section \ref{symmetry}.

A maximal set of mutually orthogonal minimal enclosing subspaces
enables now the discussion of processes inside.

\subsection{Collection into subspaces, dissipation inside}\label{dissipation}

In studying the time evolution in \cite{BNT08} we identified the condition ``laziness''.
A subspace $P\Hh$ is ``lazy'' if there is no flow out of it in first order of time.
This condition appears here as the first part for characterizing
collecting spaces, the orthogonal complements of decaying spaces.

\begin{define}{\bf Lazy subspaces:}
$P\Hh$ is a  lazy subspace if
\beq
\forall\rho=P\rho P :\,\,\,\der \Tr[P\rho(t)P]= 0\quad \textrm{at} \quad t=0.
\eeq
\end{define}
\begin{lem}\label{lazylem}
$P\Hh$ is a lazy subspace
$\iff$
\beq\label{lazyformula}
\forall\, \alpha:\quad h_\alpha\cdot P=P\cdot h_\alpha\cdot P.
\eeq
\end{lem}
\begin{proof}
Since $\TT^t$ conserves the trace, one has $\Tr[P\dot{\rho}(t)P]= -\Tr[P^\perp\dot{\rho}(t)P^\perp]$.
At time $t=0$ this is to be expressed with the generating operator as
$$-\Tr[P^\perp\D(\rho)P^\perp]=-\sum_\alpha\Tr[P^\perp h_\alpha\rho h_\alpha^\dag P^\perp]
=-\sum_\alpha\Tr[(P^\perp\cdot h_\alpha\cdot P)\rho (P\cdot h_\alpha^\dag\cdot P^\perp)].$$
So, if $P^\perp\cdot h_\alpha\cdot P$ does not vanish for each $\alpha$, there exists a state $\rho=P \rho P$,
such that $\Tr[P\dot{\rho}(t)P]\neq 0$.
This inequality holds in particular for any $\rho=P \rho P$ which has $\rank(\rho)=\Tr[P]$.
\end{proof}
\begin{define}{\bf Collecting subspaces}
\\$P\Hh$ is a collecting subspace, if
\beq\label{alltimes}
\forall \,t>0\quad\forall \rho : \quad\TT^t(P\rho P)=P\TT^t(P\rho P)P.
\eeq
\end{define}
\begin{lem}\label{colllem}
$P\Hh$ is a collecting subspace, iff it is a lazy subspace also satisfying
\beq\label{offzero}
P(iH-\half\sum_\alpha h_\alpha^\dag h_\alpha )P^\perp=0.
\eeq
\end{lem}
\begin{proof}
Since the Lindblad equation is of first order in taking the time-derivative,
the equation
\beq\label{first}
\forall \rho : \quad\D(P\rho P)=P\D(P\rho P)P
\eeq
is equivalent to (\ref{alltimes}).
The laziness condition is equivalent
to disabling, as is shown above, appearance of diagonal blocks $P^\perp\rho P^\perp$:
$$\forall\rho:\quad P^\perp\D(P\rho P)P^\perp =0,$$
and (\ref{offzero}) is equivalent to disabling appearance of off-diagonal blocks of $\rho(t)$,
$$\forall\rho:\quad P\D(P\rho P)P^\perp =0.$$
This is seen just by inserting the Lindblad equation (\ref{lindblad}):
\beq
P\D(P\rho P)P^\perp =  P\rho P(iH-\half\sum_\alpha h_\alpha^\dag h_\alpha)P^\perp
\eeq
\end{proof}
For the system there is no way (back) out of $P\Hh$.
But some part of $\rho$ may enter this subspace,
if $P$ does not commute with the Hamiltonian $H$ and all the transition operators $h_\alpha$.
In general $P\cdot\TT^t(P^\perp\rho P^\perp)\cdot P\neq 0$.
If $P$ does not commute with all the $h_\alpha$, this occurs for some $\rho$, for example
$\rho=P^\perp/\Tr[P^\perp]$, in first order in time, i.e.
$P\cdot\D(P^\perp\rho P^\perp)\cdot P\neq 0$.
If $P$ does commute with all the $h_\alpha$, but does not commute with the Hamiltonian,
entering the subspace occurs in second order:
$P\cdot\D^2(P^\perp\rho P^\perp)\cdot P\neq 0$; see the proof of Lemma \ref{vneumann}.

If $P^\perp$ projects onto a decaying subspace, then $P\Hh$ is collecting.
This can be seen in the structuring of $\Hh$, performed in Section \ref{cascades}.
For the simple $\D=\D_h$ there was no reason for such a detailed investigation:
A collecting subspace is there just a proper eigenspace for the eigenvalue zero of $h$.
But here the structures are in general richer.
There is still an evolution going on in the collecting subspace.
It can be seen as an evolution for density matrices defined on the reduced Hilbert space $P\Hh$.
\beq\label{brevereduced}
\rho=P\rho P\quad \Rightarrow\quad \dot{\rho} =\breve{\D}(\rho)\quad
\textrm{ defined with }\{\breve{H}=P H P,\, \breve{h}_\alpha= P h_\alpha P\}.
\eeq
If the collecting subspace is minimal, i.e. does not contain a smaller collecting subspace,
the evolution inside is of the same type as in a minimal enclosure;
it is dissipative, unless $\dim(P\Hh)=1$.

The relations between characteristic subspaces and stationary states can now be
analyzed, first for one direction of implications:
\begin{lem}\label{statlem}
If $\rho$ is a stationary state and $P$ is the projector onto its range,
then $P\Hh$ is a collecting subspace or an enclosure.
\end{lem}
\begin{proof}
This proof is done by revisiting the proofs for Lemma \ref{lazylem} and Lemma \ref{colllem},
with the special extra situation $\rho=P\rho P$ and $\range(\rho)=P\Hh$.
\end{proof}


The other relations state that
inside a minimal enclosure, with no decay inside, there holds
\begin{thm}{\bf Uniqueness of minimal stationary states.}\label{uniquethm}
Let $\K$ be a subspace which is a minimal enclosure or a minimal collecting subspace,
containing no smaller enclosure or collecting subspace.
Then there exists one and only one stationary state supported by $\K$. Its density matrix has maximal rank, \quad
$\rank(\rho)=dim(\K)$.
\end{thm}

\begin{proof}
There has to exist a stationary state $\rho$, see Theorem \ref{existence},
applied to the restricted evolution inside of $\K$.
Linearity of the Lindblad equation implies that the stationary states, if more than one, form a line or a (hyper)plane.
This would include elements in the boundary of $\s$ with lower rank; but
there are \emph{no invariant states} \emph{at the boundary of} $\s(\K)$:
Any invariant state has a collecting subspace as its range (Lemma \ref{statlem}),
and it is assumed that $\K$ contains no smaller collecting subspace.
\end{proof}

\subsection{Cascades of decay}\label{cascades}

In the cases where decay occurs it might be helpful or necessary to define a structure of
the Hilbert space, analogous to the ``energy levels'' in Grotrian diagrams of atoms.

We define the \textbf{lowest level of the Cascade}, $P_0\Hh$,
as the smallest subspace of $\Hh$ which contains all minimal collecting subspaces.
The strategy of further procedure to give more structuring is as follows:
Consider the complement of the lowest level, with the same evolution, except the flow
out into the lowest level. This evolution, acting on $P_0^\perp \Hh$, is generated by
$\D_1$, formed with $P_0^\perp H P_0^\perp$ and $\{P_0^\perp h_\alpha P_0^\perp \}$.
Then the first \textbf{higher level} $P_1\Hh$ of the Cascade for $\D$
is defined as the lowest level of the Cascade for $\D_1$.
Iteration gives a series of levels $P_0\Hh$, $P_1\Hh$, \ldots $P_n\Hh$, until $\bigoplus_i P_i\Hh =\Hh$.
\textbf{Basins}  $P_{i,j}\Hh$ in the level $P_i\Hh$ are defined as \emph{minimal} collecting subspaces
for the reduced evolutions with no flow out into the lower levels.

The precise details:

\begin{proposition}\textbf{Decomposition into basins. }\label{cascadeprop}
Each level $P_i\Hh$ can be decomposed into a direct sum of mutually orthogonal
basins $P_{i,j}\Hh$. This decomposition is either unique
or unique up to some unitary equivalence, which reshuffles basins
among a set of partners with equivalent dissipations inside.
\end{proposition}
\begin{proof}
Construct the decomposition inductively and begin with the lowest level.
Consider a sum of minimal collecting mutually orthogonal subspaces $P\Hh:=\bigoplus_j P_{0,j}\Hh$,
and consider another minimal collecting subspace $P_\sigma\Hh$, not contained in $P\Hh$.
$P_\sigma\Hh$ contains a unique stationary state $\sigma$ (Theorem \ref{uniquethm}).
If $P_\sigma\Hh$ is orthogonal to $P\Hh$, define $P_{0,j+1}=P_\sigma$ and proceed inductively.
If $P_\sigma\Hh$ is neither orthogonal to $P\Hh$ nor contained in it,
consider the Lindblad equation acting on $\sigma$,
with restriction to the subspace spanned by adding $P\Hh$ and $P_\sigma\Hh$,
in the block-matrix representation as stated in the Appendix.
There $R=P\sigma P\neq 0$, $Q=P\sigma P^\perp\neq 0$, since $P_\sigma\Hh$ is not subspace of $P\Hh$
but not orthogonal to it.
Also $S=P^\perp \sigma P^\perp\neq 0$, by positivity of $\sigma$, and
$\rank(S)=\rank(P+P_\sigma)-\rank(P)$, since $P_\sigma\Hh$ is minimal.
The condition ``laziness'' on $P\Hh$, as stated in Lemma \ref{lazylem}, is $C_\alpha=P^\perp h_\alpha P=0$.
Together with the condition stated in equation (\ref{offzero}), which is
$iPHP^\perp -\half\sum_\alpha A_\alpha^\dag B_\alpha=0$
(here with the notations as defined in Appendix, Section \ref{block}),
one gets for $\dot{\sigma}=0$ the part
$$\dot{S}=-\half\sum_\alpha(B_\alpha^\dag B_\alpha S+SB_\alpha^\dag B_\alpha)=0.$$
This implies $\forall\alpha\,:\quad B_\alpha=P\breve{h}_\alpha P^\perp=0$,
where $\breve{h}_\alpha$ is the restricted transfer operator, restricted to
the subspace $P\Hh +P_\sigma\Hh$.
Together with the ``laziness'' condition this means commutation of every restricted $\breve{h}_\alpha$
with $P$, and this implies moreover, again using equation (\ref{offzero}),
that also the restricted $\breve{H}$ has to commute with $P$.
It follows that
$$P_{0,j+1}\Hh:=(P^\perp\Hh\oplus P_\sigma\Hh)\ominus(P\Hh\oplus P_\sigma\Hh)$$
is another collecting subspace, orthogonal to $P\Hh$.

Iterating this procedure, until no other collecting subspace,
no other stationary state not already contained in $P_0\Hh:=P\Hh$ is left, gives a decomposition of the lowest level.

Now we investigate the restricted evolution of the complement of $P_0\Hh$.
Consider the subspace $P^\perp_0\Hh$ and the processes generated by
$\{P^\perp_0 H P^\perp_0, P^\perp_0 h_\alpha P^\perp_0\}$.
The collecting subspaces of this evolution, with all outflow into the lowest level disabled,
give the basins of the first higher level $P_1\Hh=\bigoplus_j P_{1,j}\Hh$.
Then one iterates the disabling of the outflow of the remaining subspace,
constructing the higher levels with basins $P_{i,j}\Hh$ until nothing more of $\Hh$ remains.

The lowest level is the subspace spanned by all possible
stationary states. So its definition is unique.
By induction, the entire decomposition into levels is unique.

Now consider two different decompositions of a level, w.l.o.g. of the lowest level.
This gives exactly the situation treated above, with at least one stationary state $\sigma$
with support in a basin $P_\sigma\Hh$ which is neither orthogonal to some basin $P_{i,j}\Hh$
nor contained in it.
This is the case, iff there are stationary phase relations $P_{i,j}\sigma P_{k,\ell}$.
The discussion of such cases is postponed to the following Section \ref{dephasing},
using the Proposition \ref{phaseprop}.
\end{proof}

The characterizations of basins involves a common Schur triangulation (\cite{L69}) in block form of all
the transfer operators. The blocks $P_{i,j} h_\alpha P_{k,\ell}$ with $i>k$ are zero.
The change in time of a basin's content $P_{i,j}\rho P_{i,j}$ consists of
\begin{itemize}
\item inflow from (several) $P_{k,\ell}\Hh$ with $k>i$,
    \\generated by $\{P_{i,j}HP_{k,\ell}+P_{k,\ell} HP_{i,j},\quad P_{i,j} h_\alpha P_{k,\ell}\}$
\item outflow into (several) $P_{k,\ell}\Hh$ with $k<i$,
     \\generated by $\{P_{i,j}HP_{k,\ell}+P_{k,\ell} HP_{i,j},\quad P_{k,\ell} h_\alpha P_{i,j}\}$
\item dissipation inside the basin,
     \\generated by $\{P_{i,j}HP_{i,j},\quad P_{i,j} h_\alpha P_{i,j}\}$.
\end{itemize}

If higher basins  $P_{k,\ell}\Hh$ are empty, all contents of the basin $P_{i,j}\Hh$ will decay if $i\geq 1$.
Dissipation leads to density matrices with full rank inside the basin; then, if $i\neq 0$,
there is some outflow since
$\sum_{k<i,\ell}P_{k,\ell}\cdot(iH+\half\sum_\alpha h_\alpha^\dag h_\alpha)\cdot P_{i,j}\neq 0$.
\begin{proposition}{\bf  Emptying of higher levels. }
Only the lowest level in the cascade carries stationary states,
formed by combinations of unique states inside each single basin; eventually there may be
stationary phase relations.
The collection of the higher levels in the cascade,
\begin{center}
$\K=\bigoplus_{i\geq 1,j}P_{i,j}\Hh\quad =P_0^\perp\Hh$,\end{center}
is the maximal decaying subspace. It is completely emptied in the course of the evolution.
\end{proposition}
\begin{proof}
There are no stationary states in $\K$, by construction.
So the diagonal blocks  $P_{i,j}\rho P_{i,j}$ with $i\geq 1$ vanish. By preservation of positivity,
the off-diagonal blocks $P_{i,j}\rho P_{k,\ell}$ and  $P_{k,\ell}\rho P_{i,j}$  have to vanish also.
\end{proof}

\subsection{Dephasing and the geometry of paths}\label{dephasing}

Here we study, as $t\to\infty$, the phase relations between basins.
Knowing that basins in the upper levels of the Cascade get empty,
and phase relations involving one or two of the decaying basins have to vanish because of preservation of positivity,
it remains to study phase relations between minimal collecting subspaces.
We may restrict the system and consider only the lowest level, $P_0\Hh$, which is collecting.
In this level the time evolution is identical to the evolution
defined in equation (\ref{brevereduced}), generated by the restricted operators.
Every collecting subspace is there, in this restriction, an enclosure.
So we simplify the discussion and consider a system with can be decomposed into minimal enclosures $P_i\Hh$
without decay.
The time evolution of each block $P_i\rho P_j$ is independent of all the other blocks.
When considering the phase relations between diagonal blocks
we may therefore simplify further, and restrict the system to a space with just two basins
$\Hh=P_i\Hh\oplus P_j\Hh$.

\begin{proposition}\textbf{Stationary phase relations. }\label{phaseprop}
A stationary phase relation $P_i\rho P_j$ exists if and only if there exists a
unitary operator $U$ commuting with $H$ and each $h_\alpha$
which intertwines between the two enclosures.
\beq
UP_j=P_iU\,,\quad U^2=\one
\eeq
The stationary phase relation is unique up to a constant factor.
\end{proposition}

\begin{proof}
Assume that such an intertwiner $U$ exists. It creates a dynamical symmetry, $\D(U\rho U^\dag)=U\D(\rho)U^\dag$,
and it transforms the stationary density matrices of the enclosures into each other,
\beq\label{unitary}
U\rho_j U^\dag=\rho_i.
\eeq
The commutation with the generating operators implies stationarity of
phase relation blocks $P_i\rho P_j=r_{i,j}U\rho_j$ and $P_j\rho P_i=r_{j,i}\rho_j U$.

On the other hand, assume that some stationary block $P_i\rho P_j$ exists. Then also
$P_j\rho P_i = (P_i\rho P_j)^\dag$ is stationary.
Since the matrices for $\rho$ in diagonal blocks are of full rank (Theorem \ref{uniquethm}), the density matrices
$$\sigma_\lambda=\half(\rho_i+\lambda P_i\rho P_j +\lambda^* P_j\rho P_i +\rho_j)$$
are positive for $|\lambda|$ small.
Now at some critical value of $|\lambda|$ the state $\sigma_\lambda$
is at the boundary of $\s$, and $\rank(\sigma_\lambda)<\rank(P_i)+\rank(P_j)$.
This implies that $P_\lambda\Hh$, the range of $\sigma_\lambda$, is an enclosure (Lemma \ref{statlem}),
and that $P_\lambda\in\{H,h_\alpha\}'$.
Since $P_i\Hh$ and $P_j\Hh$ are undecomposable and $P_\lambda$ does neither
commute with $P_i$ nor with $P_j$, the only possibility for such a situation is, that the von Neumann algebra
$\{H,h_\alpha\}'$ is not abelian, and
$$\{H,h_\alpha\}'\cong \C^2\otimes P_i\Hh,$$
including a unitary intertwiner $U$ acting as in equation (\ref{unitary}).
\end{proof}

\begin{proof}{\bf Uniqueness of decomposition into basins. }
Proposition \ref{phaseprop} gives the completion for proving Proposition \ref{cascadeprop}.
The decomposition into basins is not unique, iff there exist stationary phase relations.
And such stationary phase relations exist, iff there is a unitary equivalence
as stated in Proposition \ref{phaseprop}.
\end{proof}

The geometry of paths $\{\rho(t)\}\subset\s$
is related to the eigenvalues of the superoperator $\D$.
Each path can be decomposed into a sum of at least one stationary state
and paths of self-adjoint matrices which are eigenmatrices or pairwise sums
of eigenmatrices of $\D$.

\textbf{Special} paths for self-adjoint matrices:
\bit
\item eigenvalue zero \quad \ifa \quad stationary state
\item imaginary eigenvalue \quad \ifa \quad circular path
\item negative eigenvalue \quad \ifa \quad path leading straight to zero
\item complex pairs of eigenvalues with negative real part \quad \ifa \quad
paths formed as $\gamma\sigma(t)+\gamma^*\sigma^\dag(t)$, spiraling in to zero
\eit
General paths arise as superpositions of special paths.
Geometric considerations give some implications for the eigenvalues of $\D$:
Decay processes lead straight to the boundary of $\s$, so negative eigenvalues are involved.
Eigenmatrices whose range contains some part of $P_0^\perp\Hh$ belong
to eigenvalues with negative real part.
The eigenmatrices whose support is in $P_0\Hh$ can be chosen such that their supports are
in single collecting basins or in  blocks giving phase relations between pairs of basins.
In this way one gets a complete set of eigenmatrices spanning the space of all matrices with
support in $P_0\Hh$.
Since no path can leave $\s$, there are no eigenvalues with positive real part.
Moreover we can state
\begin{lem}\label{properzeroeigenmatrices}
To the eigenvalue zero of $\D$ there exist only proper eigenmatrices.
The corresponding eigenspace is spanned by positive density matrices.
\end{lem}
\begin{proof}
Consider the stationary eigenmatrix $\sigma$. If it is not self adjoint, then also $\sigma^\dag$
is an eigenmatrix, as are the selfadjoint $\sigma+\sigma^\dag$ and $i\sigma-i\sigma^\dag$.
If $\sigma$ is selfadjoint but not positive, consider it split as $\sigma=\rho_+-\rho_-$,
both parts being positive. Since $\TT^t$ is positivity preserving, both parts separately must be
stationary.
Now assume the existence of a generalized eigenmatrix $\tau_0$,
with $\D(\tau_0)=\sigma$, $\D(\sigma)=0$.
Again one can conclude that the analog equations should hold for the adjoint matrices
and for their linear combinations. One can therefore assume $\tau_0=\tau_0^\dag$.
Integrating the evolution equation, assuming $\tau(0)=\tau_0$, gives $\tau(t)=\tau_0+t\sigma$.
Multiplying with some small $\eps$ and adding some positive $\rho$ with full rank
would give a path starting inside $\s$ but leaving it as $t$ gets large.
This is a contradiction to the preservation of positivity, so no such $\tau_0$ can exist.
\end{proof}

The eigenvalues of $\D$ lying on the imaginary axis correspond to circular paths.
Such cases can appear for phase relations, for off-diagonal blocks of $\rho$.

\begin{thm}{\bf Dephasing and non-dephasing; eigenvalues of $\D$ }\label{dephsthm}
\begin{enumerate}
\item\label{a}
For a minimal block at the diagonal, belonging to a collecting basin, $\{\sigma=P_{0,j}\sigma P_{0,j}\}$,
there exists exactly one eigenmatrix to the eigenvalue zero.
All other eigenvalues have negative real part.
\item\label{b}
For an off-diagonal block $\{\sigma=P_{0,j}\sigma P_{0,k}\}$ where there exists an intertwiner
$UP_j=P_iU$ with $U\in\{P_0 H P_0,\,P_0 h_\alpha P_0\}'$,
there exists exactly one eigenmatrix to the eigenvalue zero.
The eigenmatrix is $U\rho_j$,
where $\rho_j$ is the stationary eigenmatrix with support in $P_{0,j}\Hh$.
All other eigenvalues have negative real part.
\item\label{c}
For an off-diagonal block $\{\sigma=P_{0,j}\sigma P_{0,k}\}$ where there exists an intertwiner
$U\in\{P_0 H P_0-(E_jP_{0,j}+E_\ell P_{0,\ell}),\,P_0 h_\alpha P_0\}'$,
there exists exactly one eigenmatrix to an eigenvalue on the axis of imaginary numbers.
The eigenvalue is $i(E_\ell -E_j)$, the eigenmatrix is $U\rho_j$,
where $\rho_j$ is the stationary eigenmatrix with support in $P_{0,j}\Hh$.
All other eigenvalues have negative real part.
\item\label{d}
For an off-diagonal block $\{\sigma=P_{0,j}\sigma P_{0,k}\}$ where there is no intertwiner
as in item \ref{b} or \ref{c}
there exist only eigenvalues with negative real part.
\end{enumerate}
\end{thm}
\begin{proof}
The existence and uniqueness of an eigenmatrix to the eigenvalue zero in cases (\ref{a}), (\ref{b}),
and the nonexistence in case (\ref{d}) are stated and then proven in Theorem \ref{uniquethm}
and in Proposition \ref{phaseprop};
then Lemma \ref{properzeroeigenmatrices} states that there are no generalized eigenspaces to this eigenvalue.

It remains to examine the existence or non-existence of other eigenvalues on the imaginary axis.
The method is the same as in the proof of Theorem 17 ``No circular paths'' in \cite{BNT08}.
We switch between the Schr\"{o}dinger and the Heisenberg picture.
$\D^\dag$ has the same spectrum as $\D$. Assume the existence of an eigenvalue $\lambda=ir$
with eigenoperator $F=P_{0,\ell} F P_{0,j}$, where $r\in\R$, so
$\exp{(t\D^\dag)}F=e^{irt}F$, and use the Kadison inequality.
No further details for the time evolution are needed to deduce equation (\ref{vunitary}).
We refer to \cite{BNT08} for description of how to conclude that $V:=F/\sqrt{\|F^\dag F\|}$
is a local isometry between $P_{0,j}\Hh$ and $P_{0,\ell}\Hh$ or
a local unitary if $j=\ell$:
\beq\label{vunitary}
V^\dag\cdot V=P_{0,j},\quad V\cdot V^\dag=P_{0,\ell}.
\eeq
Now we use the evolution equation (\ref{heisenberg}) for $V$, multiply from the left by $V^\dag$, and get
\beq\label{vdagd}
V^\dag\D^\dag(V)=i(V^\dag H V-P_{0,j}H)+
\sum_\alpha (V^\dag h_\alpha^\dag V h_\alpha-\half P_{0,j} h_\alpha^\dag  h_\alpha
-\half V^\dag h_\alpha^\dag  h_\alpha V) =irP_{0,j}.
\eeq
Since, by definition of the collecting basins, $\{P_{0,j}, P_{0,\ell}\}\subset\{P_0 H P_0, P_0 h_\alpha P_0\}'$,
we may define
$$\check{h}_\alpha =P_{0,j}h_\alpha = h_\alpha P_{0,j},\quad
\hat{h}_\alpha =V^\dag \check{h}_\alpha V,\quad
\check{H} =P_{0,j}H,\quad
\hat{H}=V^\dag \check{H} V,$$
and write the trace of (\ref{vdagd}) as
\beq\label{tracevdagd}
i\Tr[\hat{H}-\check{H}]+\sum_\alpha(\Tr[\hat{h}^\dag_\alpha \check{h}_\alpha] -
\half\Tr[\check{h}^\dag_\alpha \check{h}_\alpha] -\half\Tr[\hat{h}^\dag_\alpha \hat{h}_\alpha] )=ir \dim(P_{0,j}H).
\eeq
The Cauchy-Schwarz inequality, applied to inner product
$\Tr[.\cdot .]$ in the H.S. space of matrices, and the inequality between geometric and arithmetic mean
imply that the real part of the l.h.s. of (\ref{tracevdagd}) is zero iff
$\forall \alpha\,:\quad \hat{h}_\alpha=\check{h}_\alpha$, i.e.
$V^\dag h_\alpha V=P_{0,j} h_\alpha$, $h_\alpha V=V h_\alpha$.
Using again (\ref{vdagd}) this implies $H\cdot V=V\cdot(H+r)$.

There are no such imaginary eigenvalues, no circular paths, in cases \ref{a},\ref{b},\ref{d}.
They exist only in case \ref{c}, with $U=V+V^\dag+\bigoplus_{k\neq j, k\neq\ell}P_{0,k}$.
\end{proof}

\subsection{Stationary states, collection of results}\label{statstates}

A collection of results stated above in this Section gives now the proofs of the main theorems.
\begin{proof} For {\bf Theorem \ref{structurthm} on structuring of the Hilbert space: }
In Proposition \ref{cascadeprop} on decomposition into basins
the decomposition of $\Hh$ into $P_0\Hh$, defined as the lowest level
of the cascade of decay, and its complement $P_0^\dag\Hh$ is performed.
This gives part (1) of the Theorem.

The same Proposition \ref{cascadeprop} gives also the further splitting of $P_0$ into basins.
These are minimal collecting subspaces. If there exist stationary phase relations
between two basins, and only then, can this splitting be varied, using another pair of basins
which are related to the former two and among themselves by unitary transformations.
This is stated and then proven in Proposition \ref{phaseprop} on stationary phase relations.
Their existence comes with a form of equivalence between basins,
given by the unitary transformation, which has to commute with every $P_0h_\alpha P_0$ and with $P_0 H P_0$.

There may exist a generalized form of equivalence between basins,
where the unitary transformation commutes again with every $P_0h_\alpha P_0$ but then
with $P_0 H P_0 -E_jP_{0,j}+E_\ell P_{0,\ell}$ instead of $P_0 H P_0$.
Such a generalized form of equivalence implies again equivalence of the
unique stationary states located in the basins;
and it enables the occurrence of undamped oscillating phase relations.
Collecting equivalent and generalized equivalent basins into subspaces $Q_{0,k}\Hh$
gives the larger, unique, part of splitting $P_0=\sum_k Q_{0,k}$.
There are neither stationary nor undamped oscillating phase relations $Q_{0,k}\Hh Q_{0,\ell}$.
This is stated and then proven in Theorem \ref{dephsthm} on dephasing and non-dephasing.

The recollection of basins can be seen as $Q_{0,k}\Hh=\C^n(k)\otimes\Hh_{00,k}$.
Uniqueness of the stationary states $\rho_k$ in the minimal collecting subspaces,
everyone equivalent to $\Hh_{00,k}$, is stated in Theorem \ref{uniquethm}.
So parts (2) (3) and (4) of the Theorem are proven.

The splitting of $P_0^\dag\Hh$ into higher levels and basins is
stated in Proposition \ref{phaseprop}, proving the last part of the Theorem.
\end{proof}

A subspace $\K$ is the minimal support of a minimal stationary state iff it fulfills the conditions
\begin{itemize}
\item ``Laziness'' -- the projector $P_\K$ fulfills equation (\ref{lazyformula}),

\item ``No creation of off-diagonal elements'' -- $P_\K$ fulfills equation (\ref{offzero}),

\item ``Minimality'' -- $\K$ contains no smaller subspace fulfilling the first two conditions.
\end{itemize}
The characteristic equations have been found also by B. Kraus et al., \cite{K08a,K08b},
as determining ``dark states'', which are pure stationary states.

A collection of results gives also the
\begin{proof} For {\bf Theorem \ref{conventional} on enclosures and blocks: }
The relations of projectors $Q_m\in\{\Hh ,h_\alpha\}'$
to the time evolution and to ``enclosure'' are analyzed in Lemma \ref{vneumann}
and in the further discussions in Section \ref{consevation} on ``Conservation of projectors and enclosures''.
Because of the enclosure, all further decompositions can be performed for the restricted
evolution acting on the set of density matrices $\rho$ with support in $Q_m\Hh$.
\end{proof}

\section{Invariance (conservation) and symmetry}\label{symmetry}

\subsection{Invariant observables}\label{invariance}

The stationary states span a subspace of the HS-space of matrices, the space of eigenmatrices to
the eigenvalue zero of the superoperator $\D$.
Its adjoint, $\D^\dag$, has the same set of eigenvalues, so there is a linear set of invariant
operators. It has the same dimension as the set of stationary states,
and contains only proper ``eigenmatrices'', in duality to Lemma \ref{properzeroeigenmatrices}.

Let us start a construction of an invariant observable with a nucleus,
located in a collecting basin, say $P_{0,k}\Hh$.
The defining condition (\ref{alltimes}) ``nothing goes out'' is equivalent to
the dual condition ``no observable comes in''
\beq\label{nothingin}
P_{0,k}\TT^{t\dag}(F) P_{0,k}= P_{0,k}\TT^{t\dag}( P_{0,k}F P_{0,k}) P_{0,k}.
\eeq
Inside the basin is full dissipation with only one stationary state.
There is therefore only one invariant observable inside the basin (up to constant factors).
It is the projector $P_{0,k}$.
\beq
P_{0,k}\TT^{t\dag}( P_{0,k}) P_{0,k}=P_{0,k}.
\eeq
The Heisenberg evolution goes backward.
It lets, applied to the projector $P_{0,k}$, the diagonal block $P_{0,k}$ unchanged,
it creates an extension into the decaying subspace $P_0^\perp$, and also phase relations
between this subspace and the basin where it started.
It lets the whole block $P_0\Hh$ unchanged, since this collecting subspace
is spanned by collecting basins, each one showing the ``no observable comes in'' condition (\ref{nothingin}).
The evolved observable therefore stays HS-orthogonal to eventually existing
undamped oscillating phase relations, and we can define
\beq\label{definvobs}
A_{0,k}:=\lim_{t\to\infty}\TT^{t\dag}(P_{0,k})
\eeq
as an invariant observable. It is positive, since preserving positivity
goes over from $\TT^t$ to $\TT^{t\dag}$ by duality.

For a maximal set of mutually orthogonal collecting basins one obtains
\beq
\sum_kP_{0,k}=P_0 \qquad \iff\qquad \sum_k A_{0,k}=\one .
\eeq
There may be a still larger set of linearly independent invariant observables.
Let us represent subspaces with $n$ equivalent collecting basins $P_{0,\ell}$
allowing for stationary phase relations, but not undamped oscillating ones, as
$$\bigoplus_\ell P_{0,\ell}=\C^n \otimes\Hh_{00}.$$
For any $n\times n$ matrix $M$ the observable $M\otimes\one$ can be extended,
analogously to the procedure (\ref{definvobs}), to an invariant observable.
$n^2$ of these observables can be linearly independent,
n of them are as constructed in (\ref{definvobs}).

An algebraic analysis of the set of stationary states,
of invariant observables and their relations to the generating operators relies on
first cutting off the decaying subspace $P_0^\perp\Hh$.
The restricted time evolution,
generated by $\D$ defined with Hamiltonian and transfer operators $\{P_0HP_0, P_0h_\alpha P_0\}$,
is identical to the full one for $\rho$ with support on $P_0\Hh$.
For this restricted system, on $P_0\Hh$,
the concepts of ``basin'' and ``enclosure'' are identical.
Projectors onto basins are elements of the von Neumann algebra
$\Nn_0=\{P_0HP_0, P_0h_\alpha P_0\}'$,
which is the set of invariant observables for the restricted time evolution.
The set of extended invariant observables in the large system
is thus related to the commuting von Neumann algebra of the restricted system;
restricted to the non-decaying level $P_0\Hh$.
\beq
\exists F :\quad \D^\dag(F)=\dot{F}=0\qquad \iff\qquad P_0FP_0\in \Nn_0
\eeq
So, in systems without decay, the invariant observables
do form an algebra. But, in systems with decay, the extension (\ref{definvobs})
does in general destroy this property.
See examples in Section \ref{examples}.

\subsection{Symmetries}\label{symmetries}

A dynamical symmetry is defined by existence of unitary or anti-unitary operators $V$, such that
\beq\label{defsymm}
\forall t, \,\forall\rho\quad\TT^t(V\rho V^{-1})=V\rho(t)V^{-1}.
\eeq
This is equivalent to
\beq\label{defsymm2}
\forall\rho\quad\D(V\rho V^{-1})=V\D(\rho)V^{-1}.
\eeq
Sometimes an appearance of a dynamical symmetry is in connection with existence of an
\emph{algebra} of conserved observables, a connection well known in Hamiltonian dynamics.
If $U\in\{H,h_\alpha\}^\prime$ then $\TT^t(U\rho U^\dag)=U\rho(t)U^\dag$.
Another way how a symmetry may be guaranteed is, that $V H V^{-1}=H$ and
the set $\{V h_\alpha V^{-1}\}$ equals the set $\{h_\alpha\}$.
But it may also be hidden, not immediately to
be observed in the $h_\alpha$. See examples in Section \ref{examples}.
For continuous groups, and for $\D$ with a finite number of transfer operators $h_\alpha$, the symmetry is necessarily
not completely represented by invariance of the set of generators.

The \emph{maximal} symmetry is invariance under \emph{all} unitary and anti-unitary transformations.
There is only one ray of generators compatible with this symmetry,
in the center of the {\it cone of generators}, see \cite{BNT08}. It consists of
$\{\D|\,\D(\rho)=\lambda\cdot (\omega-\rho)\}$,
where $\omega=\one/\dim(\Hh )$ is the completely mixed state.
There are several ways of choosing the set $\{h_\alpha\}$ to
form such a special $\D$.
Examples are again in Section \ref{examples}.

While, in Hamiltonian dynamics, the appearances of
\begin{itemize}
\item dynamical symmetry
\item invariant (conserved) observables
\item algebra formed by the invariant observables
\end{itemize}
must come together, these relations are not strict in irreversible dynamics.
Here we observe cases of
\begin{itemize}
\item dynamical symmetry without invariant observables,
\item Invariance of observables without a symmetry,
\item Invariant observables which do not form an algebra.
\end{itemize}
For each of these cases we present examples.

There remains a relation between a dynamical symmetry and the set of stationary states.
If $V$ is a symmetry operator as in (\ref{defsymm}), then the set of stationary states obeys
the \emph{symmetry for stationarity}
\beq\label{rhosymm}
\{\rho|\, \textrm{stationary}\}=\{V\rho V^{-1}|\,\rho\,\, \textrm{stationary} \}.
\eeq
But this is, in general, a one-way relation. The symmetry for stationarity (\ref{rhosymm}) may be valid,
without (\ref{defsymm}) being true. This comes, again, because of the
restriction of the set of invariant states to the subspace $P_0\Hh$.

\section{Examples}\label{examples}

The density matrices are representations of states in some basis
of the usual type, employing a complete orthogonal set of basis vectors.
Mostly we use $\D$ given as a sum of two simple generators, with transition operators $\{h_+, h_-\}$.
Zeroes as matrix elements are represented with dots.
Lower indices on $\C$ indicate the role of a subspace as a level or as a basin.
Matrix elements of the density operator are denoted as $r_{i,j}$.
Representations using a tensor product are in accordance to its use in part (3) of Theorem \ref{structurthm}.
``No invariant observable'' means, precisely:
Only the constants are invariant.

\subsubsection{Dissipation}\label{dissipex}
Hilbert space $\Hh=\C^2$
$$h_+ =
\left( \begin{array}{cc} \cdot&1\\ \cdot&\cdot
         \end{array}\right)\qquad
h_- =
\left( \begin{array}{cc}  \cdot&\cdot\\
         1&\cdot \end{array}\right)$$
There is a unique stationary state. It is $\omega=\one/2$, a fact which can not occur with simple generators.
To represent the dynamics we use Pauli matrices, so $h_\pm=\sigma_\pm=(\sigma_x+\sigma_y)/2$,
and $\D(\sigma_\pm)=-\sigma_\pm$, $\D(\sigma_z)=-2\sigma_z$.
The dynamics is symmetric under rotation around the z-axis, under reflection $\sigma_z\leftrightarrow-\sigma_z$
and under complex conjugation $C$.
The symmetry for stationarity is maximal, but there is no invariant observable.

\subsubsection{Decay, two collecting basins, no stationary phase relations}\label{decaynoex}
$\Hh=\C^3=\C_{0,1}\oplus\C_{0,2}\oplus\C_1$
$$h _+ =
\left( \begin{array}{ccc} 1&\cdot&1\\ \cdot&\cdot&\cdot\\
         \cdot&\cdot&\cdot \end{array}\right)\qquad
h_- =
\left( \begin{array}{ccc} 1&\cdot&-1\\ \cdot&\cdot&1\\
         \cdot&\cdot&\cdot \end{array}\right)$$
         Extremal stationary states are $1\oplus 0\oplus 0$ and $0\oplus 1\oplus 0$.
The symmetry for stationarity includes exchange of these extremal states and complex conjugation.
The whole system is only one enclosure.
Invariant observables are diagonal matrices with components $(1,0,2/3)$, or $(0,1,1/3)$,
and their linear combinations. They do not form an algebra.
No dynamical symmetry (but $C$).

\subsubsection{Decay, two collecting basins with stationary phase relations}\label{decaywithex}
$\Hh=\C^3=\C_{0,1}\oplus\C_{0,2}\oplus\C_1$
$$h_+ =
\left( \begin{array}{ccc} \cdot&\cdot&1\\ \cdot&\cdot&\cdot\\
         \cdot&\cdot&\cdot \end{array}\right)\qquad
h_- =
\left( \begin{array}{ccc} \cdot&\cdot&1\\ \cdot&\cdot&1\\
         \cdot&\cdot&\cdot \end{array}\right)$$
Stationary states: Any $2\times 2$ density matrix with support on $\C_{0,1}\oplus\C_{0,2}$.
Symmetry for stationarity: $U\oplus\one$ with any unitary $U$, and $C$.
No dynamical symmetry but $C$.
Invariant observables: Linear combinations of $A_1 \ldots A_4$,
$$A_1 =
\left( \begin{array}{ccc} 1&\cdot&\cdot\\ \cdot&\cdot&\cdot\\
         \cdot&\cdot&2/3 \end{array}\right),\quad
A_2 =
\left( \begin{array}{ccc} \cdot&\cdot&\cdot\\ \cdot&1&\cdot\\
         \cdot&\cdot&1/3 \end{array}\right),\quad
A_3 = A_4^\dag=
\left( \begin{array}{ccc} \cdot&1&\cdot\\ \cdot&\cdot&\cdot\\
         \cdot&\cdot&1/3 \end{array}\right).$$
         They do not form an algebra.

\subsubsection{Decay, basins with dissipation inside}\label{decbasd}
$\Hh=\C^4=\C_0^2\oplus\C_1^2$
$$h_\pm =
\left( \begin{array}{cccc} \cdot&\pm 1&1&\cdot\\ 1&\cdot&\cdot&\cdot\\
         \cdot&\cdot&\cdot&\pm 1\\  \cdot&\cdot&1&\cdot \end{array}\right)$$
         The restricted evolution on $\C_0^2$ is almost the same as in \ref{dissipex},
         only with twice the speed.
         One stationary state, $\omega \oplus 0$, no invariant observables.
         Symmetry for stationarity: all the unitary and anti unitary transformations acting on $\C_0^2$.
         No invariant observables.

\subsubsection{Dephasing of two enclosures}\label{dephasex}
$\Hh=\C^4=\C_{0,1}^2\oplus\C_{0,2}^2$
$$h_+ =
\left( \begin{array}{cccc} \cdot&1&\cdot&\cdot\\ \cdot&\cdot&\cdot&\cdot\\
         \cdot&\cdot&\cdot&1\\  \cdot&\cdot&\cdot&\cdot \end{array}\right)\qquad
h_- =
\left( \begin{array}{cccc} \cdot&\cdot&\cdot&\cdot\\ 1&\cdot&\cdot&\cdot\\
         \cdot&\cdot&\cdot&\cdot\\  \cdot&\cdot&-1&\cdot \end{array}\right)$$
          The restricted evolutions on each $\C_{0,j}^2$ is the same as in \ref{dissipex}.
Acting on the off diagonal blocks, these evolutions are ``out of phase'',
destroying every phase relation. Such an off diagonal block evolves according to
$$\frac d{dt}\left( \begin{array}{cc} r_{1,3}&r_{1,4}\\ r_{2,3}&r_{2,4}
         \end{array}\right)=
         \left( \begin{array}{cc} r_{2,4}-r_{1,3}&-r_{1,4}\\ -r_{2,3}&-r_{1,3}-r_{2,4}
         \end{array}\right)$$
        Two extremal stationary states $\omega \oplus 0$ and $0 \oplus \omega$,
         two invariant observables, $\one \oplus 0$ and $0 \oplus \one$, very rich symmetry.

\subsubsection{Undamped oscillating phase relation}\label{undamped}
$\Hh=\C^4=\C_{0,1}^2\oplus\C_{0,2}^2\cong\C^2\otimes\C_{0,0}^2$
$$H=\left( \begin{array}{cccc} 1&\cdot&\cdot&\cdot\\ \cdot&1&\cdot&\cdot\\
         \cdot&\cdot&\cdot&\cdot\\  \cdot&\cdot&\cdot&\cdot \end{array}\right)\qquad
h_+ =
\left( \begin{array}{cccc} \cdot&1&\cdot&\cdot\\ \cdot&\cdot&\cdot&\cdot\\
         \cdot&\cdot&\cdot&1\\  \cdot&\cdot&\cdot&\cdot \end{array}\right)\qquad
h_- =
\left( \begin{array}{cccc} \cdot&\cdot&\cdot&\cdot\\ 1&\cdot&\cdot&\cdot\\
         \cdot&\cdot&\cdot&\cdot\\  \cdot&\cdot&1&\cdot \end{array}\right)$$
         The evolutions of states of the enclosures $\C_{0,j}^2$ are again as in \ref{dissipex}.
         Here they are in phase when acting on the off diagonal blocks,
         leaving a special phase relation undamped. But the Hamiltonian creates an oscillation.
$$\frac d{dt}\left( \begin{array}{cc} r_{1,3}&r_{1,4}\\ r_{2,3}&r_{2,4}\end{array}\right)
         =-i\left( \begin{array}{cc} r_{1,3}&r_{1,4}\\ r_{2,3}&r_{2,4}\end{array}\right) +
         \left( \begin{array}{cc} r_{2,4}-r_{1,3}&-r_{1,4}\\ -r_{2,3}&r_{1,3}-r_{2,4}
         \end{array}\right).$$
$$\textrm{So }\qquad\TT^t\left( \begin{array}{cc} r_{1,3}&r_{1,4}\\ r_{2,3}&r_{2,4}\end{array}\right)
         \sim_{t\to\infty}\sim e^{-it}\,\,\frac{r_{1,3}+r_{2,4}}{2}\,\,
         \left( \begin{array}{cc} 1&0\\ 0&1\end{array}\right).$$
         The equation for full dynamics can be written as
$$\frac d{dt}(\textrm{M}\otimes\rho)=-i[H,\textrm{M}]\otimes \rho +\textrm{M}\otimes\D_{0,0}(\rho),$$
with $\D_{0,0}$ as in \ref{dissipex}.
         Stationary states are the same as in \ref{dephasex}. They can be represented as tensor products of
         diagonal $2\times 2$ matrices $\textrm{M}$ with $\omega$.
         Symmetry of stationary states includes reflection, exchanging $\rho_1$ and $\rho_2$.
         Invariant observables are given by all linear combinations of the two projectors onto $\C_{0,j}^2$.
         They form an algebra.

\subsubsection{Stationary phase relation}\label{stphrel}
The transition operators $h_\pm$ are as above in \ref{undamped}, but $\D$ is given without the Hamiltonian.
Very rich symmetry, much more symmetry operations than in \ref{dephasex}.
There are more stationary states than in \ref{dephasex},
they can be represented as tensor products $\textrm{M}\otimes\omega$ with
         any $2\times 2$ matrix $\textrm{M}>0$, $\Tr[\textrm{M}]=1$.
         Invariant observables are  $\textrm{A}\otimes\one$; they form an algebra.

\subsubsection{Cascade of decay}\label{cascadeex}
$\Hh=\C^4=\C_0\oplus\C_{1,1}\oplus\C_{1,2}\oplus\C_2$
$$h_\pm =
\left( \begin{array}{cccc} \cdot&1&\pm 1&\cdot\\ \cdot&\cdot&\cdot&\cdot\\
         \cdot&\cdot&\cdot&1\\  \cdot&\cdot&\cdot&\cdot \end{array}\right)$$
         There are two lines of flow: $|1,1\rangle\langle 1,1|\,\rightarrow|0\rangle\langle 0|$
and $|2\rangle\langle 2|\,\rightarrow|1,2\rangle\langle 1,2|\,\rightarrow
|0\rangle\langle 0|$.
Both lines have the same end.
The differential equations are $\dot{r}_{1,1}=r_{2,2}+2r_{3,3}$
$\dot{r}_{2,2}=-r_{2,2}$, $\dot{r}_{3,3}=-2r_{3,3}+r_{4,4}$,
$\dot{r}_{4,4}=-r_{4,4}$ for the diagonal matrix elements.
The off-diagonals are just exponentially decaying, with one accompanying part of the flow,
$\dot{r}_{1,3}=-2r_{1,3}+r_{2,4}$ and the same for the adjoint.
No invariant observables.

\subsubsection{Maximal symmetric evolution}
$\Hh=\C^n$;\quad The evolution is $$\dot\rho=\dim(\Hh)\cdot(\omega -\rho).$$
$\D$ can be defined with \quad $H=0$, $\{h_{i,j}=|i\rangle\langle j|\}$ \quad for some basis $\{|i\rangle\}$.
There is a unique stationary state, it is $\omega =\one /\dim(\Hh)$.
Symmetry under every unitary and anti-unitary transformation holds.
Another way to represent this evolution is to choose the Weyl operators as $\{h_{i,j}\}$ \cite{BA08}.

This is a special case of \emph{detailed balance} at infinite temperature
(see the remark around equation (2.15) in \cite{G78} and references therein),
which appears in all those cases, where $\{h^\dag_\alpha \}=\{h_\alpha\}$.
(Each $h_\alpha$ either has a dual companion
$h_\beta=h_\alpha^\dag$, or it is self-adjoint $h_\alpha^\dag=h_\alpha$.)
One consequence of these symmetries of detailed balance is the
invariance of the completely mixed state $\omega=\one/\dim(\Hh )$.
Another, related, consequence is the self-adjointness of the superoperator $\D$, if $H=0$.
The examples \ref{dissipex}, \ref{undamped} and \ref{stphrel} are cases of detailed balance.

\begin{figure}
\begin{center}
\includegraphics[width=\textwidth]{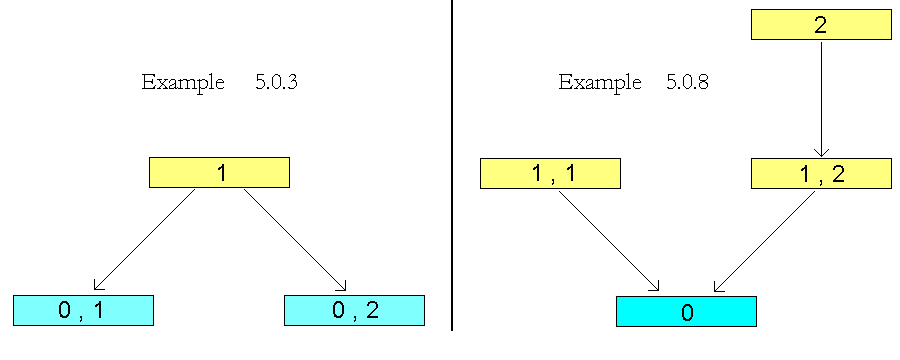}
\\{\footnotesize Figure 1. Two cascades of decay. Basins are represented by rectangles, decaying as yellow,
collecting as blue. Each flow between two basins is indicated by an arrow.\par}
\end{center}
\end{figure}

\section{Perturbations}\label{perturbation}

\subsection{Abstract considerations}\label{abstractpert}

We consider generators $\D$ depending on a parameter $\lambda$.
The dependence on $\lda$ of the transition operators is assumed as linear, which makes a
quadratic dependence on $\lda$ for $\D_\lda$.
We set $H_\lda=H_0+\lda V+\lda^2W$, and $h_{\alpha,\lda}=h_\alpha+\lda k_\alpha$.
This gives $\D_\lda=\D_0 +\lda\E +\lda^2 \F$ with
\begin{eqnarray}
\D_0(\rho)&=&-i[H,\rho]+\sum_\alpha \{
h_\alpha\rho h_\alpha^\dag -\half(h_\alpha^\dag h_\alpha\rho+\rho h_\alpha^\dag h_\alpha)\}, \\
\E(\rho)&=&-i[V,\rho]+\sum_\alpha \{h_\alpha\rho k_\alpha^\dag+ k_\alpha\rho h_\alpha^\dag \nonumber\\
& & -\half(h_\alpha^\dag k_\alpha\rho+\rho h_\alpha^\dag k_\alpha
+k_\alpha^\dag h_\alpha\rho+\rho k_\alpha^\dag h_\alpha)\}, \label{supere}\\
\F(\rho)&=&-i[W,\rho]+\sum_\alpha \{
k_\alpha\rho k_\alpha^\dag -\half(k_\alpha^\dag k_\alpha\rho+\rho k_\alpha^\dag k_\alpha)\}. \label{superf}
\end{eqnarray}
This formalism includes cases where the set of transition operators is enlarged.
Formally, it is done by perturbing some $h_\alpha=0$ with $k_\alpha\neq 0$.

A general fact is that the defining equations for \emph{structuring},
i.e. (\ref{lazyformula}) for ``laziness'' and (\ref{offzero}) for ``nothing goes out'',
and for symmetry, i.e. (\ref{defsymm2}), can abruptly turn to inequalities
through an infinitesimal change $\lda\rightarrow\lda+d\lda$,
but not the other way round:
\begin{proposition}\label{cont}
Consider the projection operators $P(\lda)$ onto subspaces, which are either
enclosures, or collecting subspaces, or basins in the cascade of decay.
Consider also the unitary and antiunitary operators $V(\lda)$ which are symmetry operations
as defined in Section \ref{symmetries}.
The functions $\lda \mapsto P(\lda)$ and $\lda \mapsto V(\lda)$ are continuous functions,
defined on closed sets of $\lda\in\R$.
\end{proposition}
\begin{proof}
To study ``laziness'' under the influence of perturbations $h_\alpha \mapsto h_\alpha(\lda)$
define $f(P,\lda):=\sum_\alpha \|h_\alpha(\lda)P-Ph_\alpha(\lda)P\|$ for orthogonal projectors $P$.
The projectors $P$ can be expressed by a finite set of parameters, e.g. its matrix elements. The function
$f(P,\lda)$ is jointly continuous in $P$ and $\lda$, so the set $\{(\lda,P)|\,f(P,\lda)=0,\,P=P^2=P^\dag\}$,
which defines lazy subspaces via the implicit functions $P(\lda)$, is closed.
It is compact, when $\lda$ is restricted to a compact interval.
The implicit functions $P(\lda)$ can be multi-valued; they give minimal lazy subspaces and direct sums of them.
The domain of $\lda\in\R$ for one $P(\lda)$ is closed, the projector
may disappear under infinitesimal changes of $\lda$.

To study ``collecting'' subspaces, consider the functions
$f_C(P,\lda):=\|P(iH(\lda)-\half \sum_\alpha h_\alpha^\dag(\lda) h_\alpha(\lda))P^\perp\|$,
and proceed in the same way as for laziness.
To study symmetry, use $f_S(V,\lda):=\|\D_\lda(V\rho V^{-1})-VD_\lda(\rho )V^{-1}\|$.

\end{proof}
Infinitesimal changes of the transition operators
and of the Hamiltonian can lead to mergers, may disturb an existing structure, or move it in the Hilbert space,
but they can not create a new one.
Moving basins in Hilbert space can not occur through mere addition of new transition operators,
since the condition (\ref{lazyformula}) for laziness involves each single $h_\alpha$.

A list of things that can happen:
\begin{itemize}
\item The number of zero eigenvalues may decrease (but it can not increase).
Example: Disturb \ref{decaynoex} with $k_\pm=\pm |0,1\rangle\langle 0,2|$.
\item Stationary phase relations may turn to undamped oscillating ones, or they may become unstable.
     The inverse changes are not possible.
Example: Disturb \ref{decaywithex} with the Hamiltonian $H=\pm |0,1\rangle\langle 0,1|$ or
with $k_\pm=\pm |0,1\rangle\langle 0,1|$.
\item Oscillating phase relations may disappear.
Example: Disturb \ref{undamped} with the $h_\pm$ of \ref{dephasex}.
\item Enclosures may merge.
Example : Disturb \ref{dephasex} with $k_+=|0\rangle\langle 1|\otimes\one$.
\item Collecting basins may merge. Example: The same as for the first item.
\item Collecting basins may merge with decaying basins.
Example: Disturb \ref{cascadeex} with $k_\pm=\pm |0\rangle\langle 1,2|$.
\item Rotations of basins in Hilbert space.
Example: Consider the simple generator with transition operator $h=\sigma_x$, disturb it with $k=\sigma_y$.
\item Symmetries may disappear (but new symmetries can not emerge).
Example: Below, in \ref{onlyone}, the worked out perturbation of \ref{dissipex}.
\end{itemize}

The general perturbation theory of linear operators, \cite{Kato}, on finite dimensional spaces
states the analyticity of eigenvalues and eigenprojectors onto eigenspaces, with only algebraic singularities
at some exceptional points.
Now we are interested in the eigenvalue zero. Its multiplicity may decrease under perturbation,
but the eigenvalue zero has to remain, with multiplicity one at least.
We want to follow this eigenvalue and the remaining eigenprojectors onto the stationary states.
The constant function $\lambda\mapsto 0$ is obviously analytic.
So we have analyticity of the eigenprojector, from which we pick out the projectors
onto those eigenmatrices which are states.
Relying on the analyticity at $\lda=0$ for
$$\lda\,\mapsto\,\{H_\lambda, h_{\alpha,\lda} \}\,\,\mapsto\,\,\\D_\lda\,\,\mapsto\,\,\{\rho(\lambda)\}$$
we make an ansatz, expanding stationary states:
\beq
\rho(\lda)=\rho+\sum_{n=1}^\infty \lda^n\sigma_n .
\eeq
We demand $\Tr[\rho(\lda)]=1$, which gives
\beq
\Tr[\rho]=1,\qquad \Tr[\sigma_n]=0.
\eeq
Expansion of the eigenvalue equation $\D_\lda (\rho(\lda))=0$ gives
the starting condition
\beq
\D_0(\rho)=0,\label{eigenrho}
\eeq
and a series of equations to determine the following contributions
\beq\label{iteration}
\D_0(\sigma_n)=-\E(\sigma_{n-1})-\F(\sigma_{n-2}),
\eeq
using $\sigma_{-1}=0$ and $\sigma_0=\rho$.
Trying to solve one of these equations one encounters two problems:
Solving it requires that the r.h.s. is a matrix in $\range[\D_0]$.
Note that the range of all the operators involved in (\ref{iteration})
consists of matrices with trace zero. This is part of solution to this first problem.
The second problem: an inverse of $\D_0$ is not uniquely given, if (\ref{eigenrho}) allows for more than one solution,
i.e. if more than one stationary state exists.
Choosing the right solution appears in the iterated equations
as solving there the first problem, deciding
whether the r.h.s. of (\ref{iteration}) is in the range of $\D_0$.

In order to proceed with calculations we define an inverse of $\D_0$ independently of the other super-operators:
\beq
\range[\D_0^{-1}]:=\range[\D_0]=\,\{\tau|\,\,\forall A_i\,\,\textrm{which are invariant:} \,\, \Tr[A_i\tau]=0\},
\eeq
with using a maximal set of linearly independent invariant operators $A_i$.
This makes the operator $\D_0^{-1}$ unique:
If there exists $\sigma$, such that $\D_0(\sigma)=\tau$, then also $\D_0(\sigma+ \sum_j\alpha_j\rho_j)=\tau$,
with HS-orthogonal eigenmatrices -- stationary states and stationary phase relations -- $\rho_j$.
Choose the $A_i$ in such a way that 
$\Tr[A_i\rho_j]=\delta_{i,j}$, then
\beq
\Tr[A_i(\sigma+\sum_j\alpha_j\rho_j)]=\Tr[A_i\sigma]+\alpha_i=0
\eeq
determines the $\alpha_i$ uniquely.

With this generally defined inverse of $\D_0$ one can invert (\ref{iteration}) to
\beq
\sigma_n=-\D_0^{-1}(\E(\sigma_{n-1})+\F(\sigma_{n-2}))+ \sum_j\alpha_j\rho_j,
\qquad \textrm{with }\sum_j\alpha_j\Tr[\rho_j]=0,
\eeq
and the task is now to find the right coefficients $\alpha_j$,
so that the insertion of $\sigma_n$ into the next iteration formulas gives
matrices in the domain of definition of $\D_0^{-1}$ -- which is the range of $\D_0$.
Here we refrain from establishing a complete formalism,
we represent several case studies instead, including special formulas.

\subsection{Case studies and examples}

\subsubsection{Only one stationary state for the unperturbed system}\label{onlyone}
If there is only one $\rho$ as eigenmatrix to the zero-eigenvalue of $\D_0$,
there are no more tasks to fulfill than to perform the calculations.
It is not necessary to make a difference between systems with decay and those without.
One may define superoperators $\G_n$ as
$$\G_0=\one,\qquad \G_1=-\D_0^{-1}\circ\E,\qquad
\G_n=-\D_0^{-1}\circ(\E\circ\G_{n-1}+\F\circ\G_{n-2}),$$
and use them to calculate
\beq
\rho(\lda)=\rho+\sum_n\lda^n\G_n(\rho).
\eeq
As an example to demonstrate the validity of this procedure we disturb \ref{dissipex}
with $k_+=(\one-\sigma_z)/2$, leaving $h_-$ undisturbed.
This gives $\sigma_{2n}=0$ and $\sigma_{2n+1}=(-2)^{-n}\sigma_x/2$.
The series converges for $|\lda|<\surd 2$ and can be summed up, giving
\beq
\rho(\lda)=\omega+\frac\lda{1+\lda^2/2}\sigma_x /2.
\eeq

\subsubsection{Remaining enclosures, each one with only one stationary state}

Consider the case of projectors $Q_j$ commuting with $H_\alpha(\lda)$ and with every $h_\alpha(\lda)$.
Inside each enclosure $Q_j\Hh$ the situation is the same as above, in \ref{onlyone}.
If there exist enclosures which allow pairwise stationary phase relations, collect them
as $\C^n\otimes \Hh$. The evolution takes place only in $\Hh$ and, if also the perturbation shows this symmetry,
it can again be treated as in \ref{onlyone}.
If it disturbs this symmetry, the phase relations vanish by dephasing.

We remark that besides the families of stationary states $\sum_j \beta_j\rho_j(\lda)$ found in this way
there could, formally, also be families with the $\beta_j$ depending on $\lda$.

\subsubsection{Merging of enclosures through direct dissipation}

Consider $\Hh=\Hh_1\oplus\Hh_2$, where each $\Hh_j=Q_j\Hh$ is an enclosure supporting just one stationary state $\rho_j$,
and where no stationary phase relations exist.
Perturb a (virtual) $h_\alpha=0$ with $k$, so the $\E$ in (\ref{supere}) is zero.
The perturbing transition operator connects the enclosures via $Q_1kQ_2\neq 0$,
the other off-diagonal block of the matrix $k$ may be zero or not.
We demand moreover
\beq\label{transitionpert}
Q_1k\rho_2\neq 0.
\eeq

The starting condition $\rho=\alpha\rho_1+(1-\alpha)\rho_2$, and the first of the conditions (\ref{iteration}),
which is $\F(\rho)\in\range(\D_0)$, here to be expressed as $\Tr[Q_1\F(\rho)]=0$, give,
using $Q_1=\one-Q_2$ and $\Tr[\F(\rho_1)]=0$,
\beqa
\alpha\Tr[Q_1\F(\rho_1)]+(1-\alpha)\Tr[Q_1\F(\rho_2)]&=&\nonumber\\
\Tr[Q_1k\rho_2k^\dag Q_1]-\alpha(\Tr[Q_2k\rho_1k^\dag Q_2]+\Tr[Q_1k\rho_2k^\dag Q_1])&=&0 .
\eeqa
That determines $\alpha\in (0,1]$ uniquely,
since (\ref{transitionpert}) implies $\Tr[Q_1k\rho_2k^\dag Q_1]>0$.

To solve the following conditions of (\ref{iteration}),
$$\F(\sigma_{{2n+2}})=-\F(\D_0^{-1}(\F(\sigma_{2n})))+\alpha\cdot\F(\rho_1-\rho_2)\in\range(\D_0),$$
we define the functional
$$\alpha[\sigma]=\frac{\Tr[Q_1\cdot\F(\D_0^{-1}(\F(\sigma)))]}{\Tr[Q_2k\rho_1k^\dag Q_2]+\Tr[Q_1k\rho_2k^\dag Q_1]}.$$
The sequence $\sigma_n$ is now fixed as $\sigma_{2n+1}=0$
and $$\sigma_{2n+2}=-D_0^{-1}(\F(\sigma_{2n}))+(\rho_1-\rho_2)\cdot\alpha[\sigma_{2n}].$$

\subsubsection{Dephasing perturbed by a Hamiltonian}

Consider again $\Hh=\Hh_1\oplus\Hh_2$, where each $\Hh_j=Q_j\Hh$ is an enclosure supporting just one stationary state $\rho_j$,
and where no stationary phase relations exist.
Perturb $\D_0$ with a Hamiltonian $V$, so the $\F$ in (\ref{superf}) is zero.
The perturbing $V$ connects the enclosures via $Q_1 VQ_2$ and $Q_2 VQ_1$; we demand
\beq\label{transitionpertham}
Q_1V\rho_2\neq 0.
\eeq
For simplicity, assume $Q_1VQ_1=0$ and also $Q_2VQ_2=0$.

The condition (\ref{transitionpertham}) implies,
as can be checked by considering matrix elements involving eigenvectors of the $\rho_j$,
$$[V,\rho_1-\rho_2]\neq 0.$$

Since $\D_0$ does not mix the matrix blocks $Q_j\sigma Q_k$ and
does not annihilate phase relations,
the range of $\D_0$ contains every $Q_1\sigma Q_2$, every $Q_2\sigma Q_1$,
and the inverse $\D_0^{-1}$ does exist for every off-diagonal block.
This applies to $\E(\rho)$, where $\rho=\alpha_0\rho_1+(1-\alpha_0)\rho_2$.
So the first order of perturbation theory gives no restriction on $\alpha_0$.
We proceed with
$$\sigma_1= -\D_0^{-1}(\E(\rho_2))-\alpha_0\D_0^{-1}(\E(\rho_1-\rho_2))+\alpha_1(\rho_1-\rho_2).$$
The condition $\E(\sigma_1)\in\range(\D_0)$ does not involve $\alpha_1$ but it determines the
right value for $\alpha_0$, by demanding
$$\alpha_0\Tr[Q_1\E(\D_0^{-1}(\E(\rho_1-\rho_2)))]=-\Tr[Q_1\E(\D_0^{-1}(\E(\rho_2)))],$$
in case $\Tr[Q_1\cdot\E(\D_0^{-1}(\E(\rho_1-\rho_2)))]\neq 0$.
The abstract considerations of Section \ref{abstractpert} imply that this factor has to be non-negative.
In fact, the part due to $-\rho_2$ may, with some tricky methods, be written as $\sum_\alpha\Tr[A_\alpha A^\dag_\alpha]$,
with $A_\alpha=h_\alpha\D_0^{-1}(V\rho_2)\rho_2^{-1/2}P_\rho-\D_0^{-1}(V\rho_2)\rho_2^{-1}P_\rho h_\alpha\rho_2^{1/2}$,
where $P_\rho$ is the projector onto the range of $\rho$,
and the $h_\alpha$ are the transfer operators appearing in $\D_0$.
The part of the factor involving $\rho_1$ is zero if $\rho_1 V=0$, but in general it is also non-negative.
We assume now, that at least one $A_\alpha\neq 0$, or, for short, just that the factor of $\alpha_0$,
which appears also as factor for all the following $\alpha_n$, is not zero.

The expansion of $\rho(\lda)$ proceeds with
$$\sigma_{n+1}= \D_0^{-1}(\E(\D_0^{-1}(\E(\sigma_{n-1}))))-\alpha_n\D_0^{-1}(\E(\rho_1-\rho_2))+\alpha_{n+1}(\rho_1-\rho_2).$$
The condition $\E(\sigma_{n+1})\in\range(\D_0)$ can be fulfilled by choosing $\alpha_n=\alpha[\sigma_{n-1}]$,
with the functional
$$\alpha[\sigma]=\frac{-\Tr[Q_1\cdot\E(\D_0^{-1}(\E(\D_0^{-1}(\E(\sigma)))))]}{\Tr[Q_1\cdot\E(\D_0^{-1}(\E(\rho_1-\rho_2)))]}.$$

\subsubsection{Basins of a cascade which merge by dissipation}

Consider $\Hh=\Hh_1\oplus\Hh_2$, where each $\Hh_j=Q_j\Hh$ is spanned by the $N+1$ lowest levels
$\psi_{j,0}\ldots\psi_{j,N}$ of a harmonic
oscillator. The decay is performed through the annihilation operators $a_j$ acting as transition operators.
This system is now perturbed by adding $\F$ made of three new $\D$.
Two of them involve the creation operators $a^\dag_j$,
the third one creates a dissipation between the two levels $\psi_{j,N}$.
Finding the right constants $\alpha_{2n}$ to do the expansion
$$\sigma_{2n+2}=-\D_0^{-1}(\F(\sigma_{2n}))+\alpha_{2n+2}\cdot(\rho_1-\rho_2)$$
requires now an $N$-fold iteration of applying the super-operator $\G:=-\D_0^{-1}\circ\F$.
It takes $N$ steps of applying $\G$ until the unperturbed ground states $\rho_j$ are lifted to the
$N^{th}$ level, and one more action of $\F$ to dissipate over to the other $\Hh_j$.
The procedure to get $\alpha_{2n+2}=\alpha[\sigma_{2n}]$ now involves the functional
$$\alpha[\sigma]=\frac{\Tr[Q_1\cdot(\F\circ \G^{N+1})(\sigma)]}{\Tr[Q_1\cdot(\F\circ \G^N)(\rho_1-\rho_2)]}.$$

\begin{figure}
\begin{center}
\includegraphics[width=\textwidth]{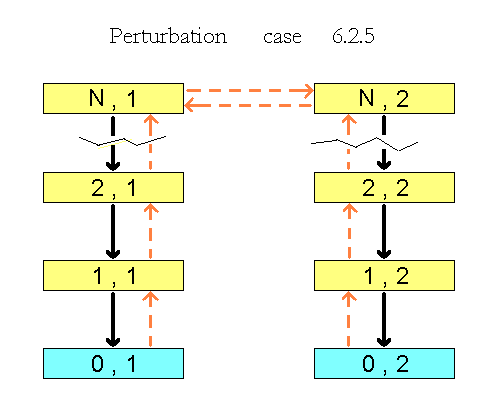}
\\{\footnotesize Figure 2. A perturbed cascade. Two independently decaying systems are perturbed
in a way as is indicated by the broken pointed lines in red.}
\end{center}
\end{figure}

\section{Summary}

Analysis of Lindblad generators is to be done on two levels.
The lower level is the finite dimensional Hilbert space $\Hh$ spanned by the pure-state-vectors.
The generators and their processes act at the upper level
which contains the mixed states represented by density matrices.
In this paper we have established an interplay of these two levels,
where every process corresponds to a structure of $\Hh$,
a decomposition into mutual orthogonal subspaces.
These subspaces, we call them basins, generalize the notion of energy levels
which appears in Hamiltonian dynamics. Their dimensions may, in general, be
any number between $1$ and the dimension of the whole Hilbert space.
They are mutually orthogonal, in spite of the possible non-hermiticity of
the super-operators which represent the Lindblad generators
and of the transfer operators acting on $\Hh$.

The interplay between process and Hilbert space structure elucidates the characterization
of the process, and brings about a way to structure the process itself,
according to its action on blocks of density matrices.
\emph{Decay} corresponds to certain subspaces which do not carry any stationary state.
Minimal stationary states are supported by minimal collecting basins,
inside of which \emph{Dissipation} occurs.
The phase relations between two basins, off-diagonal blocks of density matrices,
either show \emph{Dephasing}, or, in course of the process, go over to a special phase relation,
which is either stationary or oscillating.

Stationary and oscillating phase relations appear together with special
dynamical \emph{symmetries}. The appearances of symmetries and of invariant observables
may show peculiar effects. In cases of Decay the invariant observables
need not form an algebra. Dynamical symmetries are not the same
as the symmetries appearing for the set of stationary states, in general.
\emph{Perturbation} of a process may lead to a merging of basins,
which complicates the perturbative calculations of stationary states.

Establishing the structure connected with a process and its Lindblad generator
is, we think, a helpful tool for deeper investigations, probably indispensable.

\section{Appendix}\label{block}

\subsection*{A block matrix form of operators to characterize subspaces}
The subspace which is to be characterized, is represented with the projector
$P=\left(  \begin{array}{cc} \one & 0 \\ 0 & 0 \end{array} \right)$.
It carries the density matrices
$\rho=\left(  \begin{array}{cc} R & 0 \\ 0 & 0 \end{array} \right)$.\\
The subspace $P\Hh$ has the property of being\\
\emph{Lazy} \ifa \,
$h_\alpha=\left(  \begin{array}{cc} A_\alpha & B_\alpha \\ 0 & D_\alpha \end{array} \right).$
\\
\emph{Collecting} \ifa \, Lazy and,\, with $H=\left(  \begin{array}{cc} H_P & G \\ G^\dag & L \end{array} \right)$,
\,\, $iG -\half\sum_\alpha A_\alpha^\dag B_\alpha =0$.
\\
an \emph{Enclosure} \ifa\, \quad $H=\left(  \begin{array}{cc} H_P & 0 \\ 0 & L \end{array} \right),$
\quad $h_\alpha=\left(  \begin{array}{cc} A_\alpha & 0 \\ 0 & D_\alpha \end{array} \right).$

See also \cite{BNT08}, where formulas for calculating $\D_h(\rho)$ in block matrix form are presented.

\end{document}


\subsubsection{Decay}
$\Hh=\C^3=\C_0\oplus\C_1^2$
$$h_+ =
\left( \begin{array}{ccc} \cdot&1&\cdot\\ \cdot&\cdot&\cdot\\
         \cdot&\cdot&\cdot \end{array}\right)\qquad
h_- =
\left( \begin{array}{ccc} \cdot&i&1\\ \cdot&\cdot&\cdot\\
         \cdot&\cdot&\cdot \end{array}\right)$$
         The stationary state is $1\oplus 0$.
No dynamical symmetry, no invariant observables.

There is actually a deeper sense in the Euclidean structure thus given to $\s$;
the relations of this geometry to the algebra of physical processes is given
in Appendix \ref{algebraicgeometry}.

\subsection{Algebraic geometry of the set of states}\label{algebraicgeometry}
Euclidean geometry, drawing with ruler and compasses,
is algebraically the action of linear translations and rotations.
This can be interpreted as acting in a vector space, equipped with orthogonal transformations.
More abstract, there is the space of states = space of linear functionals on the space
of observables, equipped with unitary transformations.

Euclidean distance is additive on straight lines and invariant under rotations.
These conditions determine $d(\sigma,\rho)=\|\sigma-\rho\|$ $=(\Tr[(\sigma-\rho) ^2 ])^{\half}$.

\ldots\ldots\ldots\ldots\ldots\ldots\ldots

\subsection{Engineering}
Construction of evolutions leading to a given state,
for $\rho$ either a pure state or a state with a density matrix of full rank,
but non constant, has been presented in \cite{BNT08}.

\tab In other cases, $\rho=\omega$, or $\rho$ with rank larger than one but not full,
one has to use several transfer operators $h_\alpha$.

\begin{eqnarray}
\D_0(\rho)&=&0,\label{eigenrho}\\
\E(\rho)+\D_0(\sigma_1)&=&0,\nonumber\\
\F(\rho)+\E(\sigma_1)+\D_0(\sigma_2)&=&0,\nonumber\\
\F(\sigma_n)+\E(\sigma_{n+1})+\D_0(\sigma_{n+2})&=&0 .\nonumber
\end{eqnarray}

,
and $f_n(\lda):=\min_{P,\Tr[P]=n}f(P,\lda)$